\documentclass[runningheads,a4paper]{llncs}

\usepackage{amssymb}
\usepackage{graphicx}

\usepackage{url}
\urldef{\mailsa}\path|{luis.seoane, ricard.sole}@upf.edu|    
\newcommand{\keywords}[1]{\par\addvspace\baselineskip
\noindent\keywordname\enspace\ignorespaces#1}

\begin{document}

\mainmatter  

\title{Multiobjective Optimization and Phase Transitions}
\titlerunning{Multiobjective optimization and phase transitions}

\author{Lu\'is F Seoane$^{1,2}$ and Ricard Sol\'e$^{1,2,3}$}
\authorrunning{Lu\'is F Seoane and Ricard Sol\'e}

\institute{
  $^1$ ICREA-Complex Systems Lab, Universitat Pompeu Fabra -- PRBB, Dr. Aiguader 88, 08003 Barcelona, Spain.\\
  $^2$ Institut de Biologia Evolutiva, CSIC-UPF, Pg. Maritim de la Barceloneta 37, 08003 Barcelona.\\
  $^3$ Santa Fe Institute, 1399 Hyde Park Road, Santa Fe NM 87501, USA. \\
  \mailsa}

%
%

\maketitle

\begin{abstract}

  Many complex systems obey to optimality conditions that are usually not simple. Conflicting traits
often interact making a Multi Objective Optimization (MOO) approach necessary. Recent MOO research
on complex systems report about the Pareto front (optimal designs implementing the best trade-off)
in a qualitative manner. Meanwhile, research on traditional Simple Objective Optimization (SOO)
often finds phase transitions and critical points. We summarize a robust framework that accounts for
phase transitions located through SOO techniques and indicates what MOO features resolutely lead to
phase transitions. These appear determined by the shape of the Pareto front, which at the same time
is deeply related to the thermodynamic Gibbs surface. Indeed, thermodynamics can be written as an
MOO from where its phase transitions can be parsimoniously derived; suggesting that the similarities
between transitions in MOO-SOO and Statistical Mechanics go beyond mere coincidence.


\end{abstract}

  \section{Introduction}
    \label{sec:1}

    Optimization has always been a major topic in complex systems research. Optimality conditions
are relevant for a wealth of biological \cite{WestEnquist1999}, \cite{PerezEscuderoPolavieja2007},
\cite{ShovalAlon2012}, \cite{SchuetzSauer2012}, \cite{SzekelyAlon2013} and other natural and
synthetic systems \cite{Zipf1949}, \cite{MaritanRodriguezIturbe1996}, \cite{FerrerSole2003},
\cite{Barthelemy2011}, \cite{LoufBarthelemy2013}. Evolution through natural selection is a main
driver of biological systems towards optimal designs \cite{Dawkins1986} \cite{Dennett1995} and
certain physical principles (e.g. maximum entropy or optimal diffusion structures) already introduce
a bias towards functional extrema. Human-made systems are equally constrained through 
cost-efficiency calculations -- e.g. in transportation networks \cite{Barthelemy2011},
\cite{LoufBarthelemy2013}.

    Describing these situations requires optimal designs that often cope with interacting
constraints. To give a good account of these selective forces, a {\em Pareto} or {\em Multi
Objective Optimization} (MOO) approach can be useful. Let us introduce this theory through a recent
relevant example \cite{SeoaneSole2015a}. (Technical definitions follow below.) Consider the set
($\Gamma$) of all connected networks with a fixed number of nodes ($\gamma \in \Gamma$, figure
\ref{fig:1}{\bf a}). Among them we seek those minimizing the average path length $\left<l\right>
(\gamma)$ and the number of edges $\rho(\gamma)$. These are the {\em target functions} ($T_f(\gamma)
\equiv \{t_1 = \left<l\right>(\gamma), t_2 = \rho(\gamma)\}$) of our MOO problem. A fully connected
clique minimizes the average path length, but we need to implement all possibles links, which is
costly. The minimum spanning tree has the least number of edges possible but its average path length
is quite large. Take networks $\gamma_1$ and $\gamma_2$ trading between these extremes and such that
$\left<l\right>(\gamma_1) < \left<l\right>(\gamma_2)$ and $\rho(\gamma_1) < \rho(\gamma_1)$. This
means that $\gamma_1$ implements a better tradeoff than $\gamma_2$ and we say then that $\gamma_1$
dominates $\gamma_2$ (figure \ref{fig:1}{\bf c}). A network $(\gamma_\pi \in \Pi) \subset \Gamma$
not dominated by any other $\gamma \in \Gamma$ is {\em Pareto optimal}. Often we cannot choose
between a pair of networks because one is better than the other with respect to a target and worst
with respect to the other -- i.e. they are mutually not dominated. Because this situation is common,
MOO solutions are often not a single global optimizer, but the collection of Pareto optimal
(mutually non- dominated) networks that implement the most optimal tradeoff possible. We name this
{\em Pareto optimal set} $\Pi \subset \Gamma$.

    Target functions map each network $\gamma \in \Gamma$ into a point of $\mathbb{R}^2$:
$(\left<l\right> (\gamma), \rho(\gamma))$. This plane, with the relevant traits in its axes,
constitutes a {\em morphospace} in which salient network topologies are located as a function of
their morphology \cite{AvenaSporns2015} (figure \ref{fig:1}{\bf b}). Morphospace of other systems
visualize phenotypes or designs with respect to relevant properties. The Pareto optimal set is
mapped onto $T_f(\Pi)$ and constitutes a boundary of the morphospace (figure \ref{fig:1}{\bf b-c}).
also known as the Pareto front. 

    Some authors are beginning to explore the consequences of Pareto optimality in biological
systems \cite{ShovalAlon2012}, \cite{SchuetzSauer2012}, \cite{SzekelyAlon2013} or in relevant models
such as networks \cite{GoniSporns2013}, \cite{PriesterPeixoto2014} or regulatory circuits
\cite{SzekelyAlon2013}, \cite{OteroMurasBanga2014}. While they tackle relevant questions through MOO
methods, the description of these optimal designs is often a qualitative account of the elements
along the Pareto front (as in the study of a restricted morphospace -- an interesting contribution
nevertheless). The same qualitative bias appears in classic MOO literature. Is a more quantitative
analysis possible? Are there {\em universal features} that reach through different MOO problems,
thus uniting Pareto optimal systems despite their differences? Through our research
\cite{SeoaneSole2013} (sketched in section \ref{sec:2}) we have found a connection between MOO and
statistical mechanics. Those {\em universal features} we were looking for are phase transitions and
critical points, which leave clear imprints in the shape of the Pareto front. Some authors had
explored MOO with Single Objective Optimization (SOO) methods -- e.g. by integrating all targets
linearly to define a {\em global energy function} $\Omega(\Lambda) = \sum_k \lambda_k t_k$, with
$\Lambda = \{\lambda_1, \dots, \lambda_k\}$ arbitrary parameters that introduce a bias towards some
of the targets. Such research often finds phase transitions and other phenomena
\cite{FerrerSole2003}, \cite{LoufBarthelemy2013}. A parsimonious theory lacked as to why some
systems would present such transitions and others would not.

    To the best of our knowledge, authors researching MOO do not exploit this connection with
thermodynamics which, we believe, much enriches the discussion of Pareto optimal designs. Two
relevant examples from network theory: The efficiency of different topologies has been researched
for the relay of information across a network using two distinct delivery heuristics
\cite{GoniSporns2013}. It was made an exhaustive work in describing network topologies and locating
them in a morphospace in which different network features are segregated. In that same morphospace,
the Pareto fronts in \cite{GoniSporns2013} strongly indicate the presence of first and second order
phase transitions. If the information theoretical aspect of the diffusion of messages across the
network is considered, those transitions might become thermodynamically relevant. Similarly, the
tradeoff of topological robustness when random or targeted nodes are taken away results in a Pareto
front \cite{PriesterPeixoto2014}. Under the light of our findings, second order transitions are
present show up in that study. Also a first order transition exists that vanishes as the average
degree of the network changes, suggesting a critical point. We further illustrate our findings with
other two examples in section \ref{sec:2.01}.

    A theory about phase transitions must fit within thermodynamics. For us, this is achieved due to
the equivalence between the Pareto front and the Gibbs surface \cite{SeoaneSole2013},
\cite{Gibbs1873}, \cite{Maxwell1904}, an object known to embody phase transitions in its concavities
and non-analyticities. We discuss thermodynamics in section \ref{sec:3.01}, not because our theory
modifies previous knowledge about it, of course, but because in showing that phase transitions arise
in thermodynamics {\em precisely in the same way} as in MOO, we place our findings for MOO on very
solid ground.

  \section{Theoretical Framework}
    \label{sec:2}

        \begin{figure}
          \begin{center}
            \includegraphics[width=0.8\textwidth]{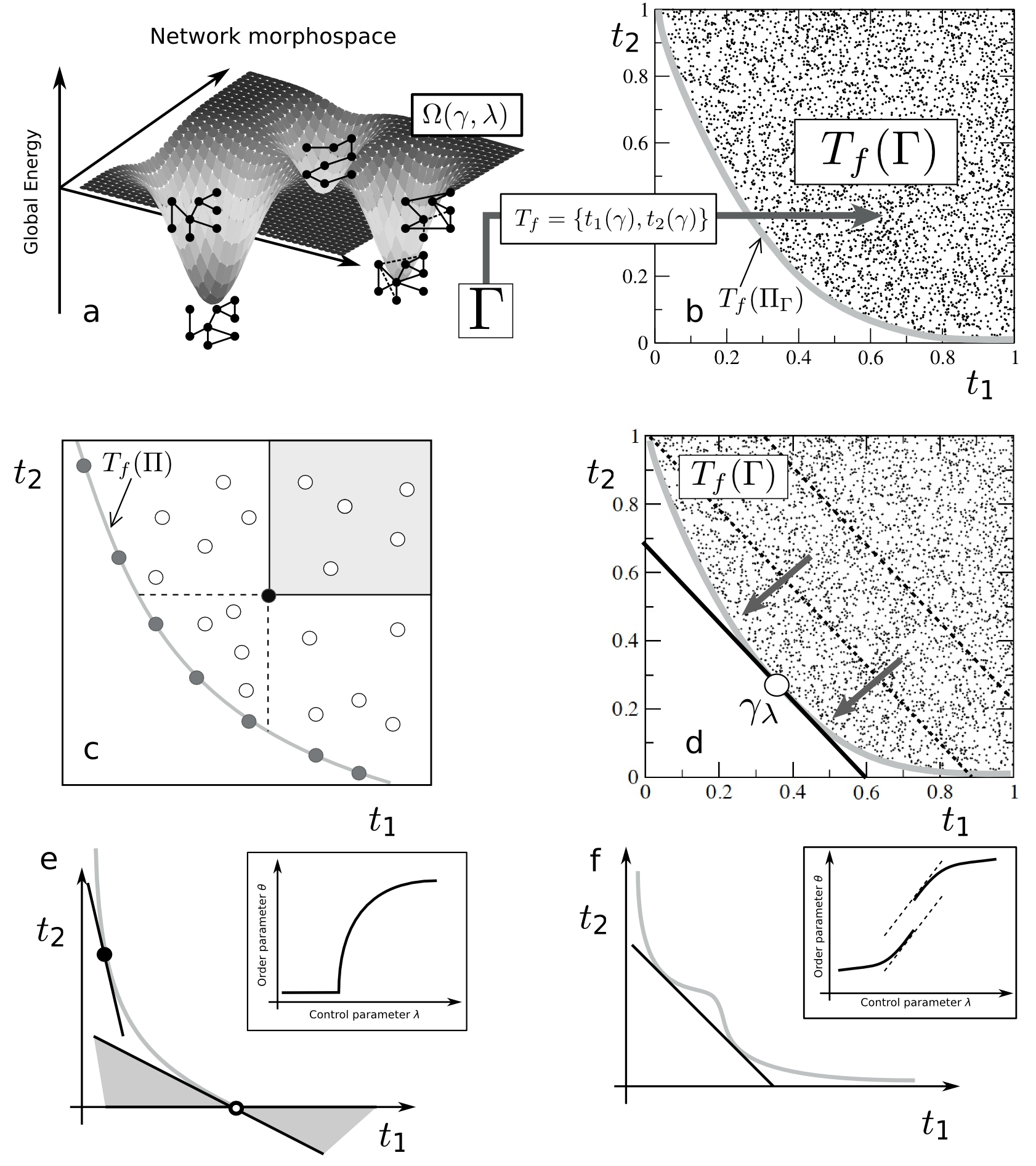}

            \caption{{\bf Phase transitions in Pareto optimal systems.} {\bf a} In a design space of
complex networks a set of weighted target functions defines a global energy (equation \ref{eq:2.01})
and renders a potential landscape (explored in \cite{SeoaneSole2013}, \cite{SeoaneSole2015a}). {\bf
b} Those same targets map the design space into the target space. The set of Pareto optimal designs
is mapped onto a boundary of this morphoscpace: the Pareto front, which represents the most optimal
tradeoff between the targets. {\bf c} The concept of dominance is geometrically simple in target
space. {\bf d} Energy minimization for fixed $\lambda$ returns a single point of the Pareto front.
Changing $\lambda$ we visit different solutions. Depending on the shape of the Pareto front, second
({\bf e}) and first ({\bf f}) order phase transitions arise as a function of $\lambda$. }

            \label{fig:1}
          \end{center}
        \end{figure}

    In this section we expand the loose introduction of MOO above. More details and methods can be
found in the exhaustive literature \cite{FonsecaFleming1995}, \cite{Dittes1996}, \cite{Zitzler1999},
\cite{Coello2006}, \cite{KonakSmith2006}. We assume minimization unless indicated otherwise.
  
    Consider a set $X$ of possible designs $x \in X$. In the example above, $X = \Gamma$ is made of
network designs. This will be used again later, along with another example in which $X = A$ stands
for all possible languages $a \in A$ derived from a mathematical computational model of human
communication \cite{FerrerSole2003}. Within $X$ we seek those optimal designs $(x_\pi \in \Pi)
\subset X$ that simultaneously minimize a series of {\em target functions} ($T_f \equiv \{t_1,
\dots, t_K\}$). These $t_k \in T_f$ map each design $x \in X$ into {\em target space} ($T_f(x) =
\{t_1(x), \dots, t_K(x)\} \in \mathbb{R}^K$), a morphospace of the system under research.

    {\em Pareto dominance} is defined in this target space. Take $x, z \in X$. $x$ dominates $z$
(noted $x \prec z$) if $t_k(x) \le t_k(z)$ for all $k$ and $t_{k'}(x) < t_{k'}(z)$ for at least one
$k'$. This means that $x$ is objectively better than $z$. If given two designs ($x, y \in X$) none
dominates the other ($x \nprec y \nprec x$), we cannot chose one of them without introducing a bias
towards some of the target functions. Pareto optimality is solved by putting choices between
mutually non-dominated designs on hold.

    The {\em Pareto optimal set} $\Pi \subset X$ is such that every element $z \in X$, $z \notin
\Pi$ is dominated by some $x \in \Pi$ while any $x,y \in \Pi$ are mutually non-dominated. The
projection $T_f(\Pi)$ conforms a $(D \le K-1)$-dimensional surface in $\mathbb{R}^K$ that embodies
the most optimal tradeoff possible between the targets. Moving along the front it is impossible to
improve all targets at once: an increment in at least one $t_k$ is necessary if we wish to decrease
some other $t_{k'}$.\\

    We sketch now the basic situations of our theoretical framework that connects the Pareto front
and thermodynamics. We refer the reader to \cite{SeoaneSole2013} for a more exhaustive discussion.

    The simplest SOO problem that includes all MOO targets defines a linear {\em global energy
function}:
      \begin{eqnarray} 
        \Omega(x, \Lambda) &=& \sum_k \lambda_k t_k(x), 
        \label{eq:2.01} 
      \end{eqnarray} 
where $\Lambda \equiv \{\lambda_k; k=1,...,K\}$ are parameters that bias the optimization towards
some of the targets. We say that equation \ref{eq:2.01} has collapsed the MOO into an SOO. A set
$\Lambda$ with fixed values $\lambda_k$ defines one single SOO, thus equation \ref{eq:2.01} (with
free $\lambda_k$) produces indeed a family of SOOs whose members are parameterized through
$\Lambda$. We will study: those SOOs, the constraints that the Pareto front imposes to their
solutions, and the relationships between different SOOs of the same family. The validity of the
results holds for any positive, real set $\Lambda$. For convenience, though: i) We take $K=2$, which
simplifies the graphic representations and contains the most relevant cases. ii) We require $\sum_k
\lambda_k=1$ without loss of generality. For $K=2$ then $\lambda_1=\lambda$, $\lambda_2=1-\lambda$,
and $\Omega = \lambda t_1 + (1-\lambda)t_2$. iii) We impose $\lambda_k\ne0\>\>\forall k$, thus
$\lambda \in (0,1)$. Comments about fringe cases can be found in \cite{SeoaneSole2013}.

    As said above, for given $\lambda_k$ one fixed SOO problem is posed. Then, equation
\ref{eq:2.01} with fixed $\Omega$ defines {\em equifitness surfaces} noted $\tau_\Lambda(\Omega)$.
Each $\tau_\Lambda(\Omega)$ constitutes a $(K-1)$-dimensional hyperplane in target space. For $K=2$
these surfaces become straight lines (figure \ref{fig:1}{\bf b}):
      \begin{eqnarray}
        \tau_\lambda(\Omega) &\equiv& \left\{(t_1, t_2)|\>\>t_2 
        = {\Omega\over 1-\lambda} - {\lambda \over 1-\lambda}t_1 \right\}. 
        \label{eq:2.01.02}
      \end{eqnarray}
The slope of $\tau_\lambda(\Omega)$ along each possible direction $\hat{t}_k$ in the target space
only depends on $\lambda$ (here, $dt_2/dt_1 = -\lambda / (1 - \lambda)$). Different
$\tau_\lambda(\Omega)$ for fixed $\lambda$ are parallel to each other. The crossing of
$\tau_\lambda(\Omega)$ with each axis is proportional to $\Omega$ (from equation \ref{eq:2.01.02},
the crossings with the horizontal and vertical axes read: $\Omega / (1-\lambda)$ and $\Omega /
\lambda$). For a given SOO (constant $\lambda$), minimizing $\Omega$ means finding
$\tau_\lambda(\tilde{\Omega})$ with $\tilde{\Omega}$ the lowest value possible such that
$\tau_\lambda(\tilde{\Omega})$ still intersects the Pareto front (figure \ref{fig:1}{\bf d}). This
is equivalent to {\em pushing} the equifitness surfaces against the Pareto front as much as possible
thus lowering the crossings with the axes.

    The SOO optimum $x_\lambda \in \Pi$ lays at the point $T_f(x_\lambda)$ at which $\tau_{\lambda}
(\tilde{\Omega})$ is usually tangent to the front (figure \ref{fig:1}{\bf d}). The exceptions to
this rule are the most interesting cases. The solutions to different SOOs (defined by different
values of $\lambda$) are found in different points along the front. For $\lambda \in (0,1)$,
equifitness surfaces present a slope $-\lambda / (1 - \lambda) = d \in (-\infty, 0)$ ($d$ decreases
as $\lambda$ increases). Consider now differentially small modifications of $\lambda$. This allows
us to drift infinitesimally slow through the SOO family. We could expect that solutions between
different SOOs will change so gradually as well, but that is not always the case.

    The front in figure \ref{fig:1}{\bf d} is convex (with respect to the optimization direction
determined by $\lambda \in (0,1)$). Its slope spans the whole range $d \in (-\infty, 0)$. This
guarantees that, as we drift through $\lambda$, each different SOO problem has one characteristic
solution laying exactly where the equifitness surface is tangent to the front. We can sample the
front smoothly, thus anything that we measure on the SOO solutions (i.e. any order parameter) will
be a smooth function of $\lambda$ as well. Convex Pareto fronts whose slope span the whole range $d
\in (-\infty, 0)$ do not present any {\em accident}.

    Consider now the case in figure \ref{fig:1}{\bf e}. It represents a convex front whose slope
spans $d \in (-\infty, d^* < 0)$. For $\lambda \in (\lambda^* \equiv -d^* / (1 - d^*), 1)$, we can
pose different SOOs whose solutions lay at different points of the convex part of the front. Varying
$\lambda$ within this interval renders a smooth sample of SOO solutions. However, for $\lambda \in
(0, \lambda^*)$ we can pose different SOOs whose solution lays exactly at the same place, as
indicated by the gray fan in figure \ref{fig:1}{\bf c}. If we measure anything about the SOO
solutions, that quantity will be constant as a function of $\lambda$ for $\lambda \in (0,
\lambda^*)$ because we will persistently measure a property of the same design. That same property
will vary smoothly over $\lambda \in (\lambda^*, 1)$. At $\lambda^*$ this quantity will be
continuous but its derivative will not (figure \ref{fig:1}{\bf e}, inset), as in second order phase
transitions. In cases like this we say that the Pareto front ends abruptly at one of its extremes.
Second order transitions also happen if the slope of the front spans $d \in (d^* > -\infty, 0)$
(i.e. if the opposite end of the front terminates abruptly) or if $d \in (-\infty, d^*_-) \cup
(d^*_+, 0)$ (i.e. the front presents a sharp edge with an ill-defined derivative).

    A cavity in the front leads to first order phase transitions. At either side of the cavity in
figure \ref{fig:1}{\bf f} we find convex stretches whose points represent different solutions for
different SOO problems posed by different $\lambda \in (0, \lambda^*)$ or $\lambda \in (\lambda^*,
1)$. But right at $\lambda = \lambda^*$ (represented by the straight red line of figure
\ref{fig:1}{\bf d}) two solutions are SOO optima at the same time. This is a phase coexistence
phenomenon characteristic of first order transitions. Pareto optimal solutions laying inside the
cavity are bypassed and never get to be SOO optima. If we measure an oder parameter of the SOO
solutions as a function of $\lambda$ (figure \ref{fig:1}{\bf f}, inset), we find a gap resulting
from the abrupt shift from one convex stretch of the front to the other at $\lambda = \lambda^*$.

  \subsection{Phase Transitions in Pareto Optimal Designs}
    \label{sec:2.01}

    As examples, we choose two problems that have recently been treated from an optimization
perspective. Take Complex Networks first, which are good models of a series of natural systems such
as vascular or nervous circuits \cite{WestEnquist1999}, \cite{PerezEscuderoPolavieja2007} that might
be constrained by physical costs (available material) while seeking the efficient implementation of
biological function (e.g. distribution of nutrients). Some human made structures, such as
transportation networks \cite{Barthelemy2011}, would also benefit from optimal design.

    \begin{figure}
      \begin{center}
        \includegraphics[width=\textwidth]{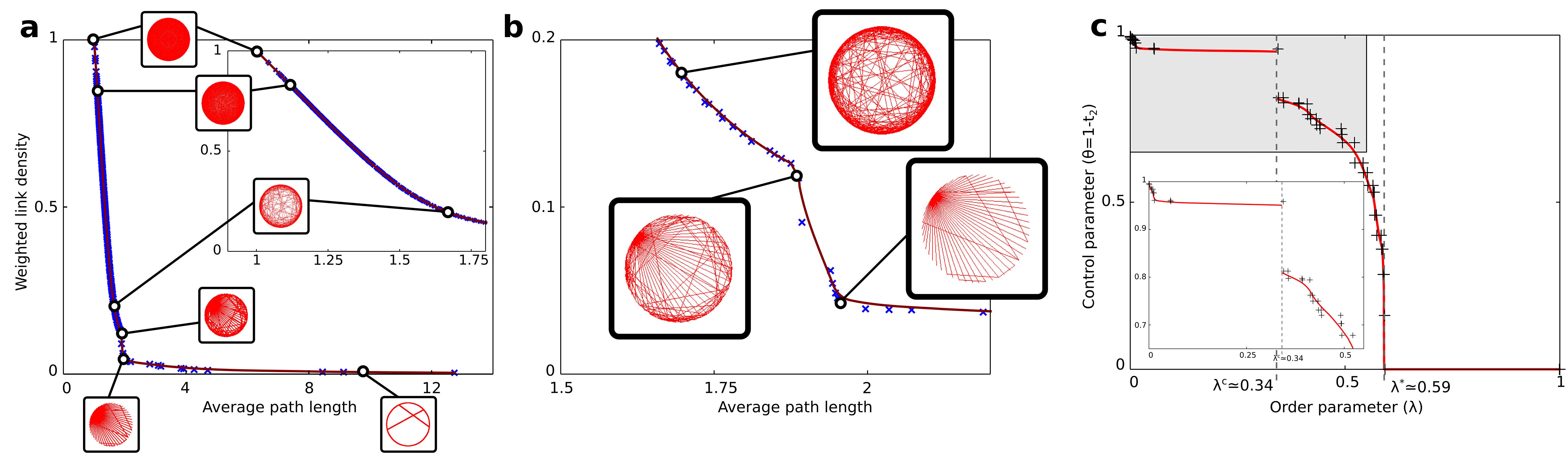}

        \caption{{\bf Pareto optimal networks with nodes spaced over a circle.} A genetic algorithm
was used to approximate the Pareto front (blue crosses and thick brown curve) of networks that
minimize the average path length and the cost of their links. {\bf a} The front implements a
tradeoff between the clique and the minimum spanning tree. (Inset) The clique extreme of the front
ends abruptly (see \cite{SeoaneSole2013}) indicating a second order phase transition. {\bf b} A
cavity is revealed at the center of the front, which implies a first order phase transition. {\bf c}
Both transitions are revealed in the plot of any order parameter. (Second order at $\lambda^*_2
\simeq 0.59$, first oder at $\lambda^*_1 \simeq 0.34$.) }

        \label{fig:2}
      \end{center}
    \end{figure}

    In \cite{SeoaneSole2015a} we consider this problem to a greater extent. We take the cost
$\rho(\gamma)$ of network $\gamma$ as a function of its edges (number or length) and its efficiency
is accounted for by the average path length $\left<l\right>(\gamma)$, a naive proxy for how fast
messages can be relayed across the network. These are the targets for minimization ($T_f=\{t_1
\equiv \rho(\gamma) , t_2 \equiv \left<l\right>(\gamma) \}$) that lead to a Pareto front and,
depending on its shape, to phase transition and other interesting phenomena. In figure
\ref{fig:2}{\bf a} we represent the front for such an MOO along with some Pareto optimal networks.
In this example the nodes are spaced over a circle and the cost of each link is proportional to its
Euclidean distance. This front ends abruptly (figure \ref{fig:2}{\bf a}, inset) and a cavity is
present (figure \ref{fig:2}{\bf b}, see \cite{SeoaneSole2015a} for discussion). This implies,
correspondingly, a second and a first order transitions at $\lambda^*_2 \simeq 0.59$ and at
$\lambda^*_1 \simeq 0.34$. These transitions can be noted in the plot of any order parameter (figure
\ref{fig:2}{\bf c}).\\

    Our second example explores the evolutionary constraints of human language, an unsettled
challenge for the scientific community. The optimization of linguistic structures brings together
universal language properties (such as Zipf's law) and the presence of ambiguity, likely as a
compromise between language economy and a large ability to talk about the outer world
\cite{SoleSeoane2014}. Such a tension was proposed by Zipf himself \cite{Zipf1949} and its
mathematical formalization \cite{FerrerSole2003} leads to an MOO problem that was always treated as
an SOO. Accordingly, phase transitions were readily identified but some debate lasted concerning its
nature and meaning \cite{ProkopenkoPolani2010}. In \cite{FerrerSole2003}, languages $a \in A$ are
modeled through a set of signals $S$ and objects $R$ whose associations are encoded in a matrix
$a=\{a_{ij}\}$ with $a_{ij}=1$ if signal $s_i \in S$ names object $r_j \in R$ and $0$ otherwise.
This binary matrix presents many ones in a row if a signal is polysemous and many ones in a column
if several words name the same object -- i.e. if there are any synonymous. Every object is recalled
equally often and, if an object has many names, a speaker chooses uniformly among these when
necessary. Two quantities are relevant (see \cite{FerrerSole2003}): i) one entropy $h_H(a)$
associated to the uncertainty of a message when a hearer has to decode it -- i.e. what object it is
meant after the speaker has uttered a signal; and ii) another entropy $h_S(a)$ associated to the
speaker choosing the right word to name an object among those available. A speaker might be allowed
to be vague (as in ``it'' referring to any object) or she might be requested to be specific (perhaps
finding the precise technicism in a scientific context).

    \begin{figure}
      \begin{center}
        \includegraphics[width=0.9\textwidth]{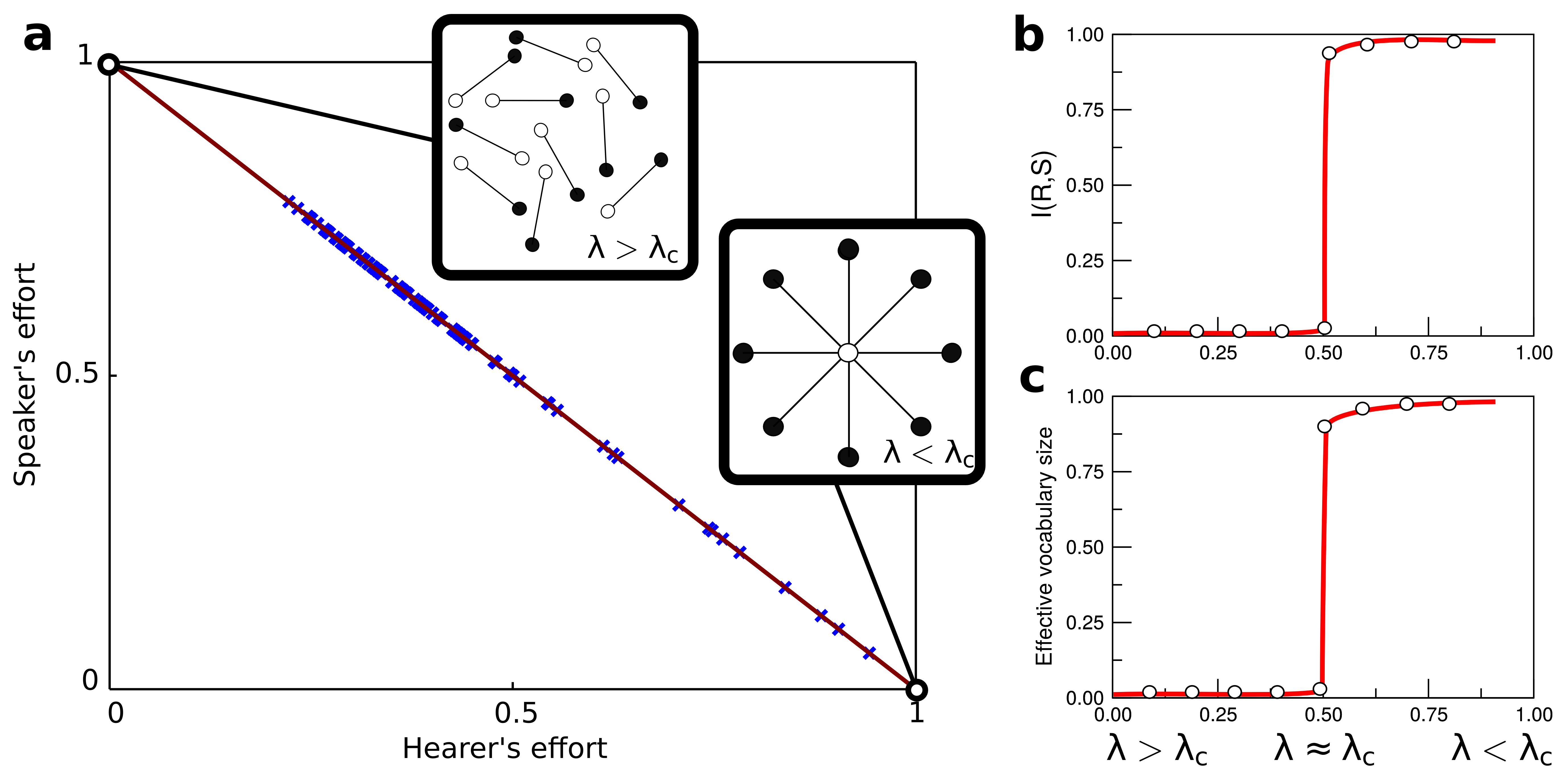}

        \caption{{\bf Least effort languages.} {\bf a} Arbitrary Pareto optimal languages (blue
crosses) lay on the straight line $t_2 = 1 - t_1$. A straight front is a sign of criticality along a
first order phase transition scenario. Either phase represents respectively the best scenario for
the speaker ($\lambda < \lambda^c$, where communication is impossible unless through the context)
and for the hearer ($\lambda > \lambda^c$, with high memory demands). Only at the critical point is
a wide complexity available. Any order parameter ({\bf b}, mutual information between the signals
and the external world; {\bf c}, effective vocabulary size) reflects the phase transition. }

        \label{fig:3}
      \end{center}
    \end{figure}

    These two entropies represent the effort made by hearers or speakers when using a language. They
act as minimization targets ($T_f=\{t_1 \equiv h_H(a), t_2 \equiv h_S(a)\}$), so that languages $a
\in A$ are subjected to a MOO. A subset $(a_\pi \in \Pi) \subset A$ of object-signal associations
implements the Pareto front, the optimal tradeoff between the efforts of a hearer and a speaker.
This front is a straight line in target space (figure \ref{fig:3}{\bf a}). Attending to the theory
sketched above this is akin to a first order phase transition (see \cite{SeoaneSole2013}) with one
end of the front being the global optimum for $\lambda < \lambda^c$ and the other extreme of the
front being optimum for $\lambda > \lambda^c$. Right at $\lambda = \lambda^c$, a sudden jump happens
between these two, very distinct phases. This can be appreciated in any order parameter as a sharp
discontinuity (figure \ref{fig:3}{\bf b}, {\bf c}). The extremes correspond to i) a language that
minimizes the speaker's efforts for $\lambda < \lambda^c$ (one single signal names every object, as
in the ``it'' example before, so that the speaker does not need to think the right association every
time) and ii) a language that minimizes the hearer's effort for $\lambda > \lambda^c$, with perfect
pairings between signals and objects so that there is not any ambiguity when decoding the messages.

    Communication is difficult in both extremes, either because the signals convey little
information about the objects ($\lambda < \lambda^c$, figure \ref{fig:3}{\bf b}), or because of the
memory needs to browse a vast vocabulary ($\lambda > \lambda^c$, figure \ref{fig:3}{\bf c}).
Besides, we know that more complicated structures exist in real languages. These structures can be
found right at $\lambda = \lambda^c$. A straight Pareto front is an indication of criticality
(\cite{SeoaneSole2013}, \cite{SeoaneSole2015b}). In such cases, exactly at the critical value
$\lambda^c$, the whole front is also SOO optimal. Note that in usual first order transitions those
solutions laying at the cavity are skipped altogether, while here a plethora of them becomes
available. In \cite{ProkopenkoPolani2010} the authors proved that the global SOO minimizers at
$\lambda = \lambda^c$ consist of all possible languages without synonyms, hence these must
constitute the Pareto front. More importantly, among these possible languages it exists one such
that the frequency of the signals obeys Zipf's law, as in natural human languages.

  \section{Discussion}
    \label{sec:3}

    \subsection{Thermodynamics as an MOO-SOO Problem}
      \label{sec:3.01}

      Thermodynamics is one of the best established branches of physics and dates back to more than
two centuries ago. In its modern form -- as statistical mechanics -- it allows us to make precise
predictions about diverse macroscopic physical phenomena. Its applications extend beyond physics, as
complex systems are increasingly being investigated through maximum entropy models \cite{Harte2011},
\cite{MoraBialek2011}. In \cite{SeoaneSole2013} we rewrite thermodynamics as an MOO-SOO problem, not
to suggest that our theoretical framework modifies it in any way. Rather the opposite: By checking
that our framework reproduces a robust physical theory, we strand our findings in a more solid
ground.

      Phase transitions in complex systems often raise heated debate: being strict, phase
transitions are defined for thermodynamics alone, through partition functions, and involve
fluctuations that compel us to take a thermodynamic limit. Little can be done against such
epistemological stand. This is yet another reason why we undertake the task of writing
thermodynamics as an MOO-SOO. Such a formalization of statistical mechanics reproduces all the
results concerning phase transitions {\em in the exact same way} that transitions arise in other
MOO-SOO scenarios. We suggest and support that the phase transition phenomenology arising in other
MOO-SOO systems is more than a qualitative similarity.

      The argument is not repeated here because of space constraints, but the idea is to show that
the {\em independent, simultaneous} minimization of internal energy and maximization of entropy
leads to a Pareto front subjected to the phenomenology found in section \ref{sec:2}. In
thermodynamics we deal with given physical systems that cannot be modified. We test probabilistic
descriptions that tell us how likely it is to find the system in each part of its phase space. We
wonder which of these descriptions present a lower internal energy and larger entropy. Thus our
design space $X$ is the set of all possible probabilistic descriptions of the system under research.
From that optimization we obtain a Pareto front whose shape (through cavities) and differential
geometry (through sharp edges) imply phase transitions if the targets ($t_1 \equiv U$, $t_2 \equiv
S$) were collapsed into an SOO problem. But that is precisely what happens in equilibrium
thermodynamics through the minimization of the free energy $F = U - TS$ \cite{Maxwell1904}. We
identify $\Omega \equiv F$, $\lambda_1 \equiv 1$, and $\lambda_2 \equiv -T$, and the theory exposed
above applies with transitions at singular temperature values.

      We insist that the optimization operates {\em upon} probabilistic descriptions of the
thermodynamic species -- while the shape of the front is determined by the properties of the
physical system. It might be interesting to segregate what thermodynamic phenomenology happens
because thermodynamic systems {\em are} probabilistic ensembles (in this regard they are unlike
Pareto optimal networks or least effort languages, as much as networks and languages are unlike each
other) and what phenomenology arises because of the shape of a Pareto front (that would yield the
same phenomenology irrespective of the kind of designs considered -- were they networks or languages
-- as long as the front had the same shape).

      This interpretation of statistic-mechanical systems is illustrated with two very simple
examples with first and second order transitions and one critical point in \cite{SeoaneSole2013}. As
stated above, this is not to prove new thermodynamic results, but to provide more solid basis for
this theory regarding MOO-SOO situations. Indeed, the role of cavities in first order phase
transitions dates back to Gibbs \cite{Gibbs1873}, \cite{Maxwell1904}, whose {\em Gibbs surface}
represented the states of a thermodynamic species. That surface is associated to the microcanonical
ensemble \cite{SeoaneSole2013} and may be concave or convex. Its convex hull is associated to the
canonical ensemble (hence to equilibrium at given temperature through free energy minimization),
which is always convex. At cavities in the Gibbs surface, the description of both ensembles must
differ (as noted by the theory of ensemble inequivalence \cite{TouchetteTurkington2004}) and first
order transitions occur.

    \subsection{Closing Remarks}
      \label{sec:3.02}

      With our recent findings \cite{SeoaneSole2013} we close a gap between the MOO literature,
research on SOO tradeoffs, and statistical mechanics. On the one hand, standard MOO analysis does
not take into account phenomena like phase transitions or criticality which, we believe, add up to
our knowledge and enrich the description of Pareto optimal designs. On the other hand, analyses of
the Pareto front are often qualitative or based on subjective appreciations of its shape. The
formalism developed in section \ref{sec:2} allows us to locate quantitatively very relevant details
of the systems under research. These features shall persist under transformations of the targets
and, if not, the qualitative description would tell us {\em how} do these phenomena disappear.
Furthermore, a solid connection to thermodynamics has been established. We are pretty confident of
the immutable, lasting nature of thermodynamics; thus we can guess that, through the Pareto
formalism, we have located broad features that unite the description of diverse MOO problems.

      A prominent field for MOO application is biology \cite{ShovalAlon2012},
\cite{SchuetzSauer2012}, \cite{SzekelyAlon2013}. Thermodynamic-like phenomenology is not discussed
in these references, but the stage looks great: Is there a place for true MOO in biology? Against
this, natural selection concerns itself with fitness maximization alone. This feels like an exciting
MOO-SOO picture, but we cannot guarantee linear global functions as in equation \ref{eq:2.01}.
Beyond linearity, new phenomenology might be uncovered.

      Finally, an important, though conceptually difficult issue was left aside in
\cite{SeoaneSole2013} and only incidentally dealt with here. How do critical systems look like under
an MOO perspective? Can we recover the astounding phenomena of criticality? This will be tackled in
future work \cite{SeoaneSole2015b}. Other theoretical aspects of MOO remain open to research.

  \subsubsection*{Acknowledgments}

    This work has been supported by an ERC Advanced Grant, the Bot\'in Foundation, by Banco
Santander through its Santander Universities Global Division and by the Santa Fe Institute. We thank
CSL members for insightful discussion.

\end{document}